\documentclass[a4,10pt]{article}
\usepackage[margin=1in]{geometry}
\usepackage{parskip}

\usepackage{titlesec}
\titleformat{\section}{\bfseries\large}{\thesection.}{0.333em}{}
\titleformat{\subsection}{\bfseries}{\thesubsection}{0.333em}{}
\titleformat{\subsubsection}{\itshape}{\thesubsubsection}{0.333em}{}
\titlespacing{\section}{0pt}{*3}{*1}
\titlespacing{\subsection}{0pt}{*2}{*0.5}
\titlespacing{\subsubsection}{0pt}{*1.5}{0pt}

\usepackage{helvet}

\everymath{\sf}
\everydisplay{\sf}

\PassOptionsToPackage{hyphens}{url}
\usepackage[colorlinks=true,
            linkcolor=blue,
            urlcolor=blue,
            citecolor=blue,
            anchorcolor=blue]{hyperref}

\usepackage[bibstyle=vancouver,citestyle=numeric-comp,doi=true]{biblatex}
\renewcommand{\cite}{\supercite}

\usepackage{caption}

\usepackage[labelformat=simple, labelfont=bf]{subcaption}

\usepackage[section]{placeins}

\usepackage{amssymb,amsmath,mathrsfs,bm,graphicx,color,float,mathtools,makecell,xparse,longtable,threeparttable,booktabs,multirow,multicol,algorithm,algpseudocode}

\usepackage{pgffor}
\newcommand{\inputtikz}[3]{%
  \includegraphics{#3.pdf}
  \foreach \l in {#2}{
  \begin{subfigure}[h]{0\textwidth}
    \phantomcaption{\label{fig:#1-\l}}
  \end{subfigure}}
}

\NewDocumentCommand{\myauthor}{m m m m}{
  \parbox{\textwidth}{
    \vspace{0.333em}
    \bfseries{#1} \mdseries\par
    ORCID:\@ #2\par
    Affiliation:\begin{itemize} \foreach \a in {#4}{\item \a} \end{itemize}
    Email: #3
    \vspace{0.333em}
  }
}

\begin{document}

\section*{Article Title:}

ByteQC: GPU-Accelerated Quantum Chemistry Package for Large-Scale Systems

\section*{Article Category:}

Software Focus

\section*{Authors:}

\begin{longtable}{|c|}
  \hline
  \myauthor{Zhen Guo*}{0000-0001-8615-3830}{gz.137@bytedance.com}{{ByteDance Research, Fangheng Fashion Center, No. 27, North 3rd Ring West Road, Haidian District, Beijing 100098, People's Republic of China}}                                                                                                                                                                                                                                                \\\hline
  \myauthor{Zigeng Huang*}{0000-0001-7963-9074}{huangzigeng@bytedance.com}{{ByteDance Research, Fangheng Fashion Center, No. 27, North 3rd Ring West Road, Haidian District, Beijing 100098, People's Republic of China}}                                                                                                                                                                                                                                       \\\hline
  \myauthor{Qiaorui Chen}{0000-0002-9830-4285}{cjerry@nvidia.com}{{NVIDIA}}                                                                                                                                                                                                                                                                                                                                                                                     \\\hline
  \myauthor{Jiang Shao}{0000-0001-9454-0558}{jiangs@nvidia.com}{{NVIDIA}}                                                                                                                                                                                                                                                                                                                                                                                       \\\hline
  \myauthor{Guangcheng Liu}{0009-0009-1993-4384}{edormancy@gmail.com}{{ByteDance Research, Fangheng Fashion Center, No. 27, North 3rd Ring West Road, Haidian District, Beijing 100098, People's Republic of China}}                                                                                                                                                                                                                                            \\\hline
  \myauthor{Hung Q. Pham}{0000-0003-3608-1298}{hung.pham@bytedance.com}{{ByteDance Research, San Jose, CA 95110, US}}                                                                                                                                                                                                                                                                                                                                              \\\hline
  \myauthor{Yifei Huang}{0000-0003-4981-7452}{huangyifei.426@bytedance.com}{{ByteDance Research, Fangheng Fashion Center, No. 27, North 3rd Ring West Road, Haidian District, Beijing 100098, People's Republic of China}}                                                                                                                                                                                                                                      \\\hline
  \myauthor{Changsu Cao}{0000-0002-2437-6825}{caochangsu@bytedance.com}{{ByteDance Research, Fangheng Fashion Center, No. 27, North 3rd Ring West Road, Haidian District, Beijing 100098, People's Republic of China}}                                                                                                                                                                                                                                          \\\hline
  \myauthor{Ji Chen}{0000-0003-1603-1963}{ji.chen@pku.edu.cn}{{School of Physics, Peking University, Beijing 100871, People's Republic of China}, {Interdisciplinary Institute of Light-Element Quantum Materials and Research Center for Light-Element Advanced Materials, Peking University, Beijing 100871, People's Republic of China}, {Frontiers Science Center for Nano-Optoelectronics, Peking University, Beijing 100871, People's Republic of China}} \\\hline
  \myauthor{Dingshun Lv*}{0000-0003-0573-6490}{lvdingshun@bytedance.com}{{ByteDance Research, Fangheng Fashion Center, No. 27, North 3rd Ring West Road, Haidian District, Beijing 100098, People's Republic of China}}                                                                                                                                                                                                                                         \\\hline
\end{longtable}

\section*{Conflicts of interest}

The authors declare that they have no conflict of interest.

\section*{Abstract}

Applying quantum chemistry algorithms to large-scale systems requires substantial computational resources scaled with the system size and the desired accuracy.
To address this, ByteQC, a fully-functional and efficient package for large-scale quantum chemistry simulations, has been open-sourced at \url{https://github.com/bytedance/byteqc}, leveraging recent advances in computational power and many-body algorithms.

Regarding computational power, several standard algorithms are efficiently implemented on modern GPUs, ranging from mean-field calculations (Hartree-Fock and density functional theory) to post-Hartree-Fock methods such as Møller-Plesset perturbation theory, random phase approximation, coupled cluster methods, and quantum Monte Carlo methods.
For the algorithmic approach, we also employ a quantum embedding method, which significantly expands the tractable system size while preserving high accuracy at the gold-standard level.

All these features have been systematically benchmarked.
For standalone algorithms, the benchmark results demonstrate up to a 60$\times$ speedup when compared to 100-core CPUs.
Additionally, the tractable system sizes have been significantly expanded: 1\,610 orbitals for coupled cluster with single and double excitations (1\,380 orbitals with perturbative triple excitations), 11\,040 orbitals for Møller-Plesset perturbation theory of second order, 37\,120 orbitals for mean-field calculations under open boundary conditions, and over 100\,000 orbitals for periodic boundary conditions.
For the advanced quantum embedding feature, two representative examples are demonstrated: the water cluster problem (2\,752 orbitals) and a water monomer adsorbed on a boron nitride surface (3\,929 orbitals), achieving the gold-standard accuracy.

With these efforts, ByteQC is expected to significantly advance research in quantum chemistry, particularly in large-scale, high-accuracy calculations.

\section*{Graphical/Visual Abstract and Caption}

\begin{figure}
  \centering
  \inputtikz{visual}{}{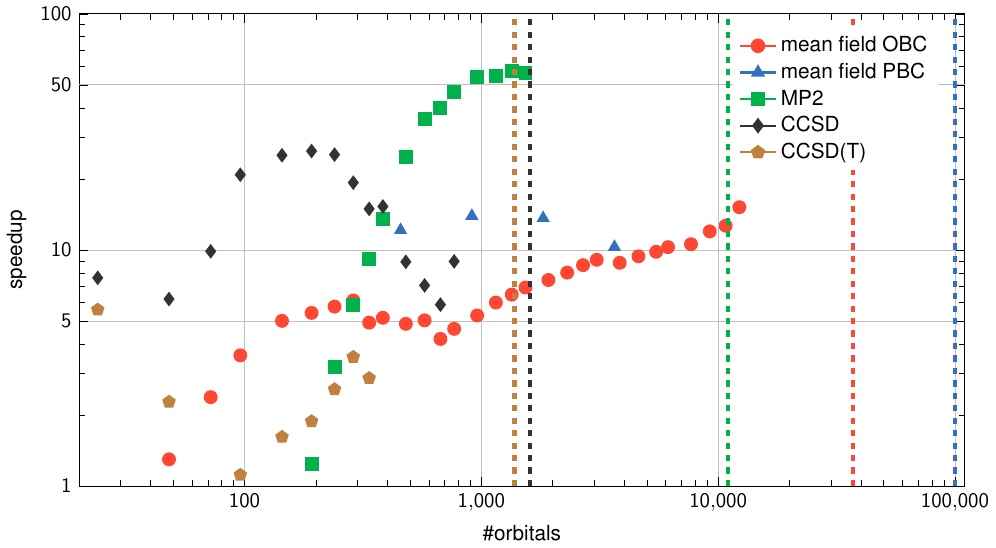}
  \caption*{\textbf{Caption}: The speedup of an NVIDIA A100 GPU compared to a 100-core CPU for individual components of the ByteQC package, using the cc-pVDZ basis set, across varying numbers of orbitals.
    The vertical dashed lines indicate the largest systems that can be handled: 1\,380 orbitals for CCSD(T), 1\,610 orbitals for CCSD, 11\,040 orbitals for MP2, 37\,120 orbitals for mean-field calculations under open boundary conditions~(OBC), and over 100\,000 orbitals (determined by the number of k-mesh points multiplied by the number of orbitals) for mean-field calculations under periodic boundary conditions~(PBC).}
\end{figure}

\newpage

\section{\MakeUppercase{Introduction}}
Quantum chemistry aims to simulate chemical systems using the principles of quantum mechanics.
Such simulations are playing an increasingly significant role in the field of drug design~\cite{Raha2007,Zhou2010,Oldfield2002,Cavalli2006,Ginex2024}, catalyst reactions~\cite{Bell2011,Kozuch2006,Jung2003,Hwang1996,Yu2013,Cortright2001,Ahn2019,Noodleman2004}, materials discovery~\cite{Brunin2019,Curtarolo2013,Chen2015,Luo2020,Curtarolo2012,Ludwig2019,Sinnott2013,Bereau2016,Cheng2014}, and so on.
An exact simulation tends to scale exponentially with system size, significantly limiting the range of systems that can be explored.
To overcome this challenge, many methods have been proposed to balance accuracy with the system size they can accommodate.
For example, methods range from those capable of handling large systems, like mean-field calculations such as density functional theory~(DFT)~\cite{Hohenberg1964, Kohn1965, Kohn1999}, Hartree-Fock~(HF) method~\cite{Hartree1935,Fock1930}, to more accurate but only limited to small to modest system methods, which include Møller-Plesset~(MP) perturbation theory~\cite{Moeller1934}, random phase approximation~(RPA)~\cite{Bohm1953,Pines1952,Bohm1951}, coupled cluster theory~(CC)~\cite{Bartlett2007}, quantum Monte Carlo method~\cite{Foulkes2001, Huang2024a}, and the density matrix renormalization group method~\cite{White1993}, among others.
However, problems that remain unresolved and are of particular interest to us, such as catalytic systems or strongly correlated systems, remain largely beyond the reach of the methods mentioned above, necessitating enhancements either in their capability to handle larger and more complex systems or in their accuracy.

In this work, we undertake efforts from two perspectives.
First, the advancements in computational power brought about by GPUs are fully utilized.
With the continuous improvement of the GPU Python ecosystem, exemplified by PyTorch~\cite{Paszke2019}, CuPy~\cite{Okuta2017}, and JAX~\cite{Bradbury2018}, GPUs have gained increasing attention for their applications in quantum chemistry simulations.
Several notable projects have already emerged, including GPU4PySCF~\cite{Li2024,Wu2024}, TeraChem~\cite{Seritan2020,Seritan2020a}, and QUICK~\cite{Miao2015,Manathunga2021,Manathunga2023}.
However, these packages do not fully meet the demands of our research, particularly in terms of the system sizes that can be computed on GPUs.
While GPUs excel at accelerating tasks requiring massive parallelism, such as linear algebra operations, which are common in quantum chemistry and often act as bottlenecks, not all quantum chemistry algorithms are naturally parallelizable.
The intricate logic of these algorithms poses significant challenges for efficient GPU implementation.
These challenges define the core objectives behind this work\@: enabling large-scale system computations within the constraints of GPU memory and efficiently implementing the complex logic of quantum chemistry algorithms on GPUs.
Second, for the algorithmic approach, we employ the quantum embedding method, particularly the density matrix embedding theory~(DMET) and its variants, such as the systematically improvable embedding~(SIE) method.
These quantum embedding methods can achieve the same level of accuracy as high-level solvers, but at a much lower computational cost, thanks to the fragment-based nature of the embedding.
Pioneer CPU-based projects have already been developed in this direction, including LibDMET~\cite{Cui2023,Zhu2019,Cui2019} and Vayesta~\cite{Nusspickel2022,Nusspickel2023}, and largely expand the range of systems that can be explored.
However, as demonstrated in this work, a fully GPU-accelerated quantum embedding method can further expand the tractable system size or enhance accuracy to the gold standard coupled cluster with single and double excitations plus perturbative triple excitations (CCSD(T)) level, using currently affordable hardware.

Combing the above advancement from both computation power and many-body method, we introduce our GPU-accelerated quantum chemistry package, ByteDance Quantum Chemistry~(ByteQC) to achieve the large-scale, efficient and accurate quantum chemistry calculation.
To begin with, we have extended existing GPU implementations or developed new GPU-accelerated versions of the canonical quantum chemistry algorithms, including HF/DFT calculations, Møller-Plesset perturbation theory of second order~(MP2), coupled cluster with single and double excitations~(CCSD), and CCSD(T).
Note that additional GPU-accelerated features, including RPA, tensor hyper-contraction~\cite{Hohenstein2012a,Parrish2012,Hohenstein2012} for exchange matrix construction, a more efficient version of multi-GPU CCSD, and auxiliary-field quantum Monte Carlo~\cite{Huang2024a}~(AFQMC) are currently under development and will be released upon completion.
Significant efforts have been made to optimize the efficiency of these functionalities and maximize the system sizes that our package can handle.
Multi-GPU support is already enabled in ByteQC, and ideal scaling is achieved in most cases.
For standalone algorithms, benchmark results show a speedup of up to 60$\times$ when using NVIDIA A100 80 GB GPUs compared to 100-core CPUs.
Furthermore, the system sizes that can be effectively handled have been greatly increased, with 1\,380 orbitals for CCSD(T), 1\,610 orbitals for CCSD, 11\,040 orbitals for MP2, 37\,120 orbitals for mean-field calculations under open boundary conditions~(OBC), and more than 100\,000 orbitals for mean-field calculations under periodic boundary conditions~(PBC).
Moreover, we implement the fully GPU-accelerated SIE framework~\cite{Nusspickel2022} by heavily reusing the above standalone module.
We benchmark the SIE framework with the water clusters and explore the potential application on the system of a water monomer adsorbed on the boron nitride surface.

Our package has already been open-source and can be found at \url{https://github.com/bytedance/byteqc}.
Several GPU techniques, such as fully functional tensor contractions and a multi-GPU class, which are considered beneficial for the development of other algorithms, have been abstracted into an independent subpackage with well-designed interfaces.
With these efforts, ByteQC will stand out as a feature-rich and highly efficient GPU package for quantum chemistry, making it ideal for large-scale research and the development of GPU-accelerated quantum chemistry software.

The paper is organized as follows. The methods and implementations of the individual subpackages are presented in Section~\ref{sec:tech}.
Section~\ref{sec:results} discusses the benchmark results and associated applications, highlighting the capabilities of our package.
An application example of a water monomer adsorbed on the boron nitride surface is provided in Section~\ref{sec:app}.
Finally, the conclusions and future outlook are provided in the concluding section.

\section{\MakeUppercase{Method and implementation}\label{sec:tech}}

\begin{figure}[htb!]
  \centering
  \inputtikz{pkg}{}{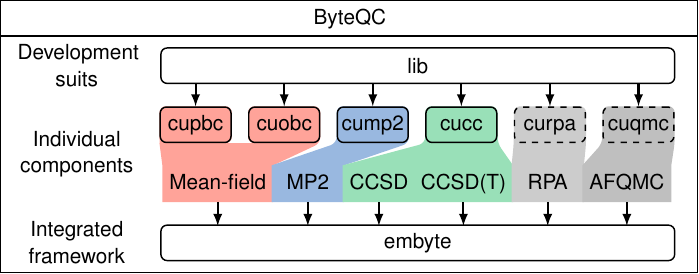}
  \caption{The software structure of ByteQC\@.
    The subpackages enclosed in dashed-line boxes are under development.
    The ``cuqmc'' subpackage is already merged into the ``ipie''  packages~\cite{Malone2022,Jiang2024,Huang2024a} and a wrapper will be included in ByteQC.\label{fig:pkg}}
\end{figure}

Currently, ByteQC provides six individual subpackages: the ``lib'' subpackage containing the common utilities useful for GPU development, the ``cupbc'' subpackage accelerating the DFT for periodic boundary condition systems, the ``cuobc'' subpackage implementing the molecules HF simulations with open boundary condition, the ``cump2'' subpackage including the GPU version of MP2 correlation energy calculation, the ``cucc'' subpackage providing the CCSD and CCSD(T) support, and the ``embyte'' subpackage integrating all the aforementioned subpackages into a SIE framework.
The other two subpackages, ``curpa'' for RPA calculations and ``cuqmc'' for AFQMC, are still under development and will be published upon completion.
The relationships between all subpackages of ByteQC are illustrated in Fig.~\ref{fig:pkg}.
All these subpackages, along with their technical details, are described in order below.

\subsection{GPU development suits}

During the development of this package, we found that many common tricks are repeatedly applied and may be reused by other quantum chemistry packages accelerated by GPUs.
We wrapped them into a standalone subpackage named ``lib'' and two of them are described in detail as follows.

First, tensor contractions are common and fundamental operations in quantum chemistry algorithms and are usually the most computationally intensive parts.
The default algorithm to perform a contraction transforms the contraction of two tensors into a general matrix multiplication~(GEMM) of two new transposed matrices and then transposes the resulting matrix into the desired tensor form, namely the transpose-transpose-GEMM-transpose algorithm.
To make GEMM efficient, transposing operations require extra memory to store the temporary results.
In the worst case, three extra memory buffers are needed to perform one contraction, which is problematic when contracting large tensors with limited memory.

Recently, some transpose-free algorithms have been proposed to perform the contraction with minimum extra memory usage and show great performance enhancements over the default one~\cite{Matthews2018}\@.
A similar implementation of transpose-free tensor contractions is available in the NVIDIA cuTENSOR library, well integrated into the Python ecosystem with CuPy.
Its multi-GPU version, cuTENSORMg, which is also part of the NVIDIA math library family, is especially well-suited for the contraction of large tensors stored in CPU memory or distributed across multiple GPUs.
The Ping-Pong scheme is automatically adopted when communicating data between the CPU and GPU, or between GPUs, to overlap communication and computation time.
Unfortunately, at the time we developed our package, there was no Python package providing the interfaces of the cuTENSORMg library.
We wrap the cuTENSORMg functions in Python and provide a user-friendly interface that has already been pushed to the main branch of the CuPy repository on GitHub.

\begin{figure}[htb!]
  \centering
  \inputtikz{con}{}{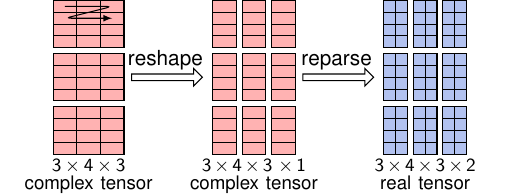}
  \caption{Reparsing a complex tensor as a real tensor with an extra size-2 axis. The black arrow indicates the linear memory layout, using col-major order as an example.\label{fig:con}}
\end{figure}

The cuTENSOR/cuTENSORMg libraries both utilize tensor cores for tensor contractions.
However, when a contraction is performed between real and complex numbers, the tensor cores are not supported according to the official documentation.
We found that by reparsing the complex tensor as a real tensor with an additional size-2 axis, as shown in Fig.~\ref{fig:con}, the contraction becomes a real tensor contraction without any data movement, enabling the tensor cores to significantly accelerate the calculation.
This provides the most efficient method for contractions between real and complex tensors.

We therefore wrap cuTENSOR, cuTENSORMg, and the complex-to-real conversion trick into one unified contraction function with the same interface, which is widely used in our package and saves a lot of development time.

Second, the multi-GPU support is another common trick in GPU development.
There are several typical methods to manage different GPUs on the CPU end.
The most straightforward approach is to launch multiple processes, each assigned to a single GPU\@.
However, dynamically launching subprocesses requires careful management of function contexts to ensure availability across all subprocesses.
The copy-on-write scheme of the fork function on Linux can simplify this, but initializing the CUDA runtime before forking prevents child processes from using CUDA functions.
Thus, we chose to use threads instead of processes.

All data and contexts are shared within threads without communication overhead, and threads can be dynamically launched without conflicting with the CUDA runtime, offering a flexible interface for extending functions to multi-GPU versions.
Note that Python's global interpreter lock prevents multiple threads from executing in parallel.
However, this is not an issue here, as launch time is usually negligible compared to the GPU running time, and the GPU kernel function runs asynchronously with CPU functions.
The only drawback of this scheme is that the developer must be careful with all CUDA memory allocations and all data transportation between CPU and GPU to achieve ideal parallelism.

All allocations from CUDA runtime, including the GPU global memory and CPU pinned memory, will lock the process-wide pagetable and will explicitly synchronize all threads, causing some threads to wait.
Data transfer from pageable CPU memory to GPU memory requires explicit allocation of pinned memory, which also causes issues.
In contrast, transfers from pinned CPU memory to GPU memory are unaffected.
With our tools, enabling multi-GPU support for a function is straightforward.
Developers can pre-allocate all CUDA memory before entering multi-GPU regions using utilities provided by our package.
After that, a single function call is sufficient to configure the number and selection of GPUs, without further modifications.
The multi-GPU related function is also available in the ``lib'' subpackage with some simple examples.

\subsection{Mean-field calculation}

The most expensive part of the mean-field calculation is the electron repulsion integrals~(ERIs) generation.
The full ERIs scale as $O(N^4)$ with increasing system size $N$, leading to significant computational and storage costs.
One trick to reduce such costs is density fitting~(DF), which utilizes an auxiliary artificial molecule to generate Cholesky-decomposition-ERIs~(CD-ERIs) with only $O(N^3)$ scaling.
The full ERIs are then obtained by contracting the CD-ERIs with themselves.
However, for large-scale calculation, $O(N^3)$ is still too large to handle for most computers, so making CD-ERIs slice-by-slice is unavoidable.
This may introduce additional overhead in constructing the exchange matrix, as the CD-ERIs are repeatedly loaded.

ERIs are only affected by the basis set and the atomic structure, remaining invariant during the self-consistent field iterations.
Thus, ERIs only need to be calculated once and stored for future use, which is the second trick to reduce the computational cost.
For nontrivial small systems, the ERIs are too large for in-core storage, requiring disk storage.
Alternatively, ERIs can be generated on-the-fly, where they are recomputed as needed.
The primary disadvantage of the on-the-fly scheme generation is its increased computational expense, especially when handling CD-ERIs slice-by-slice.
However, it eliminates the need for large disk storage and avoids data transfers between disk and CPU/GPU memory.
Time savings can only be achieved with an exceptionally efficient implementation, which is typically a significant challenge for CPU-based software.

\subsubsection{Mean-field calculation under periodic boundary conditions}

For periodic systems, ERIs further scale with the square of the number of kmesh-points, increasing the computation burden.
Thus, DF is commonly used for periodic system calculations with CD-ERIs stored on the disk.
However, as the system size grows, the required disk capacity quickly becomes unaffordable.
To alleviate this, we implement an on-the-fly version of DF and propose and implement a GPU-adapted prescreening scheme.

Although this approach may initially seem expensive, as shown below, it outperforms the disk-based scheme for building the Coulomb matrix without requiring a large disk, since the Coulomb matrix only needs two passes of CD-ERIs regeneration.
Additionally, reading from disk is serial and cannot be efficiently parallelized across multiple GPUs.
For the exchange matrix, the application of DF leads to excessive regeneration of CD-ERIs, causing significant efficiency issues.
Thus, we perform the mean-field calculations with only the Coulomb matrix and the DFT exchange matrix.
It should be mentioned that the DFT exchange matrix is usually much cheaper compared to the Coulomb matrix.
For non-trivially small systems, the wall time of DFT exchange is minimal even in CPU calculations and can be effectively overlapped with the subsequent Coulomb matrix calculation on the GPU\@.

Another common trick for periodic systems is prescreening.
If some integrals can be known to be negligibly small before computation, such as integrals of two orbitals that are far from each other in real space, skipping such integrals will save computation time.
GPUs operate ``single instruction, multiple threads'' paradigm, where minimizing branch divergence is critical for optimal performance.
A warp contains 32 threads and is the smallest instruction unit, and only one instruction can be executed among all threads at one time.
If only one thread in a warp is not prescreened, all other threads must wait for it to finish the expensive ERI calculation, negating the time-saving by prescreening.
Thus, the prescreening on GPUs requires a new design other than just following the CPU logic.~\cite{Sun2023}

A straightforward solution is to first calculate all the integral indices that are not prescreened and store them in global memory, then launch a separate kernel to read the indices and perform the integral calculation.
Since the number of prescreened integrals is unknown in advance, the pre-allocated global memory must be sufficiently large to store all indices.
Furthermore, accessing global memory is considerably slower compared to other types of GPU memory, resulting in wasted computational time as well.

\begin{figure}[htb!]
  \centering
  \inputtikz{warpspec}{pre,1,2,3,4,cal}{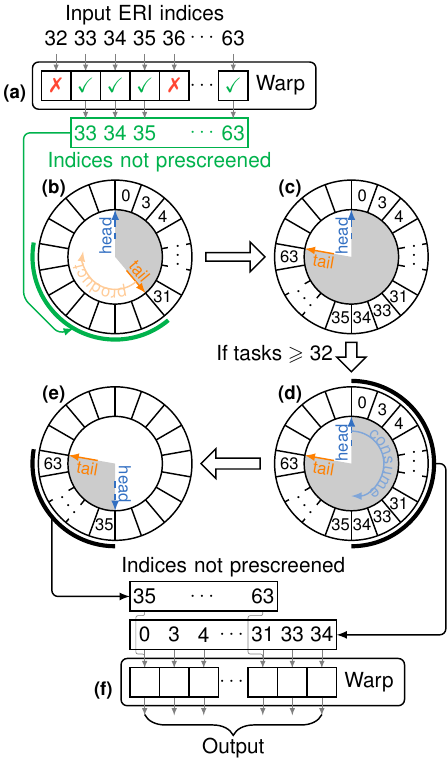}
  \caption{The diagram of the warp specialization technique used to implement the prescreening process.
    In \textbf{(a)}, the warp is initially launched to check whether the prescreening conditions are satisfied for the input ERI indices.
    Indices that pass the prescreening are placed into a loop buffer, as shown in \textbf{(b)(c)}.
    Steps \textbf{(a)-(c)} are repeated until all input ERI indices have been processed.
    During this process, whenever the number of tasks in the loop buffer exceeds 32, the first 32 tasks are removed, as depicted in \textbf{(d)(e)}, and are then processed by the warp, as shown in \textbf{(d)(f)}.
    At the end, the residual tasks will also be processed by th e warp as shown in \textbf{(e)(f)}.\label{fig:warp}}
\end{figure}

To solve this, we proposed a new scheme utilizing the warp specialization technique.
The main idea is to keep threads in a warp always in the same branch while different warps run different instructions.
The CD-ERI indices to be calculated are divided into many batches, with each batch processed by a single warp.
As shown in Fig.~\ref{fig:warpspec-pre}, all 32 threads in a warp first check whether the prescreening condition is satisfied for the input ERI indices, and place the indices that do not meet the condition into a loop-buffer stored in on-chip shared memory (Figs.\ref{fig:warp}\subref{fig:warpspec-1}\subref{fig:warpspec-4}).
Two indicators are used to mark the head and tail positions of the stored tasks in the loop buffer, as illustrated by the blue dashed arrow and the orange solid arrow in Figs.\ref{fig:warp}\subref{fig:warpspec-1}\subref{fig:warpspec-4}.
The tail indicator is incremented clockwise when tasks are added to the loop buffer, as shown in Figs.\ref{fig:warp}\subref{fig:warpspec-1}\subref{fig:warpspec-2}, while the begin indicator is incremented when tasks are removed, as shown in Figs.\ref{fig:warp}\subref{fig:warpspec-3}\subref{fig:warpspec-4}.
This mechanism prevents unnecessary copying and movement of tasks.
The process is repeated until all ERI indices have been checked.
During this iteration, whenever the number of tasks in the loop buffer reaches or exceeds 32, all threads begin to calculate the integrals and consume the first 32 tasks in the loop buffer, as shown in Figs.\ref{fig:warp}\subref{fig:warpspec-3}\subref{fig:warpspec-cal}.
The capacity of the loop buffer is 64, twice the number of threads in a warp, which is guaranteed to be sufficient, since once the number of tasks in the buffer exceeds 32, they will immediately be consumed.
With this scheme, the shared memory is reused many times to avoid storing all the tasks at once.
Besides, the shared memory is nearly two orders of magnitude faster than the global memory and greatly reduces the memory I/O time.
If all prescreening conditions are checked while there are still $n$ tasks left in the loop-buffer ($n$ is guaranteed to be less than 32), the first $n$ threads in the warp are scheduled to calculate the corresponding CD-ERIs as shown in Figs.\ref{fig:warp}\subref{fig:warpspec-4}\subref{fig:warpspec-cal}.
The pseudocode for the above scheme is presented in Algorithm~\ref{alg:warp}.
For readers unfamiliar with CUDA, the built-in CUDA function ``coalesced\_threads()'' returns a group containing only the active threads in the current warp that are currently executing a particular code path.

\renewcommand{\algorithmicrequire}{\textbf{Input:}}
\renewcommand{\algorithmicensure}{\textbf{Output:}}
\algnewcommand{\LineComment}[1]{\State \(\triangleright\) #1}

\begin{algorithm}
  \begin{multicols}{2}
    \begin{algorithmic}[1]
      \Require ERI indices to calcualte
      \Ensure ERIs
      \State new loopBuf[64] in shared memory
      \State head = tail = 0
      \For{ind in ERI indices}
      \LineComment{All threads are producers}
      \State isPrescreen = CheckPrescreen(ind)
      \If{not isPrescreen}
      \State cth = coalesced\_threads()
      \LineComment{cth: group of threads entering if branch}
      \State loopBuf[(tail+cth.rank())\%64] = ind
      \If{cth.rank() == 0}
      \State tail = (tail+cth.size())\%64
      \EndIf
      \EndIf
      \If{(tail-head)\%64 $\ge$ 32}
      \LineComment{All threads are consumers}
      \State cth = coalesced\_threads()
      \State ind = loopBuf[(head+cth.rank())\%64]
      \State CalculateERI(ind)
      \If{cth.rank() == 0}
      \State head = (head+cth.size())\%64
      \EndIf
      \EndIf
      \State res = (tail-head)\%32
      \State nth = warp.rank()
      \LineComment{Consume residual tasks}
      \If{nth $<$ res}
      \State ind = loopBuf[(head+nth)\%64]
      \State CalculateERI(ind)
      \EndIf
      \EndFor
    \end{algorithmic}
  \end{multicols}
  \caption{CUDA pseudocode of the dynamical warp specialization for prescreening.\label{alg:warp}}
\end{algorithm}

We adopt the range-separated technique~\cite{Ye2021, Bintrim2022}, and our GPU codes are derived from and benchmarked with the CPU version of the range-separated density-fitting code in repo\footnote{\url{https://github.com/hongzhouye/pyscf/tree/rsdf_direct}}.
Based on this, all core calculations are redesigned using CUDA to ensure their efficiency on GPUs.
The interfaces are also redesigned to be more user-friendly and suitable for multiple GPUs.

\subsubsection{Mean-field calculation under open boundary conditions}

For molecular systems, the ERI is computationally less expensive, and calculating the Fock matrix on-the-fly without DF is more favorable.
Additionally, CD-ERI generation is also supported in our package for other applications.

Our codes for the calculation  mean-field of molecular systems are based on the early version of the GPU4PySCF~\cite{Wu2024, Li2024}.
We made several major modifications to achieve our goal of efficient and large-scale quantum chemistry simulation:
\begin{enumerate}
  \item The early version of GPU4PySCF only supports angular momentum up to \textbf{f}.
        Higher angular momentum calculation requires numerous registers to store the temporary results, which can lead to register spilling and force the data to be stored in local memory.
        Local memory shares hardware with 80 GB global memory for A100 and is proportional to the number of the most threads the GPU can launch in the ideal case and irrelevant to how many GPU threads the kernel function actually launched.
        The required local memory for \textbf{f} orbitals is already 25.3 GB, and an out-of-memory error will occur for higher orbitals.
        We first remove the redundant temporary variables used to storage the local ERIs generated by one GPU thread, which reduce the local memory usage by more than half (this trick is also found independently in the later version of GPU4PySCF), then each high angular momentum kernel function is further divided into $N_{\text{Rys}}$ small kernels, where $N_{\text{Rys}}$ is the number of Rys roots~\cite{King1976, Rys1983}.
        Thus, the local memory required by each small kernel is further reduced by a factor of $1/N_{\text{Rys}}$ while the launched threads is enlarged by a factor of $N_{\text{Rys}}$.
        For our implementation, \textbf{f} orbitals require only 1.27 GB of local memory, and even \textbf{h} orbitals need only 6.4 GB of local memory.
        We also observed that the threads actually launched on GPU for large angular momentum kernels are usually not capable to full-fill the GPU\@.
        Thus, small kernels with more threads also lead to performance gains.
  \item For building the Coulomb and exchange matrices, the nearby $k$ threads will sum up their own results simultaneously into the global memory of GPU with the same address\@.
        Taking $k=7$ as an example in Fig.~\ref{fig:shfl}, every group of $7$ threads is colored the same.
        To avoid race conditions, atomic operations are needed here even though they will make the whole summation process serial, creating a bottleneck of the calculations.
        To minimize atomic operations, a warp-wide reduction should be first performed.
        CUDA provides two types of reduction functions.
        One function sums up the values of nearby $k$ threads, for $k\in{2,4,8,16,32}$.
        However, in our case, $k$ varies for different cases, and cannot always be guaranteed to be a power of 2.
        The other reduction function sum threads with the same labels, regardless of proximity, thus less efficient.
        In our implementation, all threads are first divided into segments based on the modulus of $k$ and their corresponding warp as shown in Fig.~\ref{fig:shfl}.
        Each thread's reversed rank $p$ within its segment is then calculated, and is indicated by the number in blocks in Fig.~\ref{fig:shfl}.
        The warp-level ``\_\_shfl\_down'' function is repeatedly called with the thread distance $\delta$ equals to $2^i$ for $i=0, 1, 2,\cdots$, until $2^i>k$.
        As a result, each thread receives a value from another thread that is $\delta$ distance away.
        However, only the thread that satisfies $p \ge \delta$ is allowed to accumulate the value, as indicated by the red solid or black dashed arrows in Fig.~\ref{fig:shfl}.
        After the iteration, the atomic operation is launched by the first thread in each segment.
        Only the additions indicated by the red arrows contribute to the final result, avoiding duplicate summation.
        This approach minimizes the number of atomic operation, accounts for thread proximity, and ensures that the performance is at least as good as the built-in ones.
        According to our tests, this scheme reduces the time required to build the Coulomb and exchange matrices by approximately half compared to the approach using atomic operations within each thread.
        \begin{figure}[htb!]
          \centering
          \inputtikz{shfl}{}{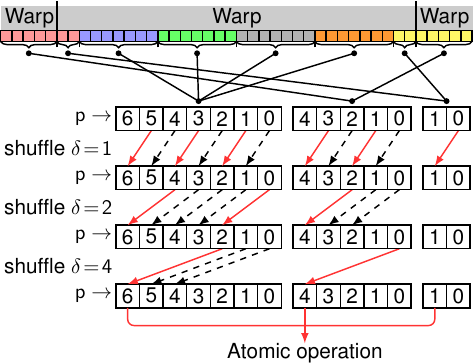}
          \caption{
            The scheme for adding the results of $k=7$ adjacent threads (with the same color) into the same memory address.
            The threads are divided into several segments based on their corresponding warp and color.
            The  ``\_\_shfl\_down'' function is called $3$ times (since $\text{ceil}(\log_2 k)=3$), with thread distances $\delta=1,2,4$.
            The thread that satisfies $p\ge\delta$ accumulates the values received from the ``\_\_shfl\_down'' functions, as indicated by the red solid and black dashed arrows, where $p$ is the thread's reversed rank within its segment.
            Finally, the atomic operation is launched by the first thread of each segment.\label{fig:shfl}}
        \end{figure}
  \item The calculations of different parts of ERIs are independent, thus parallel calculations of different parts on different GPUs will definitely reduce the wall time.
        However, achieving the ideal linear scaling with the number of GPUs is not straightforward.
        In GPU4PYSCF, some preparation calculations on CPU are needed and the results (pageable memory) are copied to GPU before launching the GPU kernel.
        As we mentioned before, the multi-GPU is achieved by multiple threads.
        The CPU calculation requires careful management of CPU resources, and the copy process also synchronizes all GPUs, both of which make multi-GPU setups less straightforward.
        We have optimized such CPU calculations, made them efficient on GPU and the copy is then naturally avoided.
        As a result, our GPU kernels are totally asynchronous with the CPU, and the ideal linear scaling verse the number of GPUs can be achieved as mentioned bellow.
\end{enumerate}

\subsection{MP2}

MP2 is a typical post-HF ab initio method in the field of computational chemistry.
The main formula of MP2 reads
\begin{equation}
  E^{\text{MP2}} = \sum \frac{(ia|jb)[2(ia|jb) - (ib|ja)]}{\varepsilon_i+\varepsilon_j - \varepsilon_a -\varepsilon_b},
\end{equation}
with $i,j$ the doubly-occupied orbitals, $a,b$ the unoccupied virtual orbitals, $(ia|jb)$ the molecular orbital ERIs, and $\varepsilon_{(i,j,a,b)}$ the corresponding molecular orbital energy.
The MP2 calculation requires only ERIs and the molecular orbital energies, but it still remains challenging for large systems.

If one treats the denominator as a rank-4 tensor $D_{ij}^{ab} = \varepsilon_i+\varepsilon_j - \varepsilon_a -\varepsilon_b$, the computation becomes straightforward, but it will require large memory to store $D_{ij}^{ab}$.
A better way is to use the CUDA kernel to read molecular orbital energies and perform the division in-place.
For systems with even 8\,000+ orbitals, storing the molecular orbital energies amounts to approximately 62.5 KB, which is well within the L1 cache capacity of A100 GPUs.
Consequently, optimizing data reading through shared memory is not particularly necessary in this case.
We utilize the friendly kernel interface ElementwiseKernel provided by the CuPy package to achieve the in-place operation.
The kernel interface of CuPy will compile the code into a shared library the first time the function is called and reuse the compiled file for later usage.
It has the same efficiency as the CUDA kernel and provides a more flexible way to accelerate some simple functions without extra CUDA file generation.

For large systems, the molecular orbital ERIs are commonly computed in advance in DF form and stored on disk.
The contractions in the above formula are too large to be handled in a single step, necessitating a slice-by-slice approach.
This requires reading the ERIs from disk multiple times.
Consequently, the time spent reading data from disk into GPU memory becomes a bottleneck.

To address this, we first implement a Ping-Pong scheme to overlap I/O operations with computations.
In our Ping-Pong scheme, the first slice is read from disk into pinned memory on the CPU and then transferred to the GPU directly.
While the first slice is being processed in the following calculations, the second slice is read and transferred concurrently.
Achieving perfect overlap here is challenging, as the computation-to-I/O time ratio varies depending on the cases and hardware.
In our hardware, the disk is distributed across the network.

To further optimize, we develop a multi-process version of the file-reading class, following the h5py package's interfaces, with an additional parameter setting to store large tensors as multiple smaller files.
When a read/write operation requires data from different files, it is automatically split into several smaller read/write operations, launched simultaneously across multiple processes, while maintaining the same interface as standard operations.
We found that launching 8 processes efficiently utilizes our network-distributed disk.
Note that these configurations are highly hardware-dependent, and for some systems, no speedup may be gained from multi-process reading.

\subsection{Coupled cluster theory}

The CC theory extends the mean-field wave functions into many-body wave functions using the exponential cluster operator to account for electron correlation.
The higher the excitations included in the cluster operator, the greater the accuracy of the wave function, but the scaling of the theory increases as well.
The CCSD method, which only considers singles and doubles excitations in the many-body wave function, has a computational scaling of $O(N^6)$ and a storage scaling of $O(N^4)$.
The perturbative energy correction upon CCSD given by partial triple excitation terms is termed as CCSD(T), which is believed to be the golden standard of quantum chemistry calculations with a higher computational scaling $O(N^7)$.

\subsubsection{CCSD}

The calculation of CCSD is composed of a series of contractions which are, in principle, suitable for GPUs and can be easily implemented by substituting the contractions with their GPU counterparts, e.g.\ from ``numpy.einsum'' to ``cupy.einsum''.
However, achieving both efficiency and support for large-scale computations in a GPU implementation of CCSD presents two main challenges.

First, the tensors involved in these contractions scale quartically with the number of orbitals, potentially exceeding GPU memory limits for systems with a large number of orbitals.
To mitigate this issue, tensor contractions are meticulously reordered to minimize the number of temporary tensors required.
In addition, a transpose-free algorithm is employed to eliminate the extra memory typically required by tensor contractions.
Custom CUDA kernel functions are also designed to perform as many in-place operations as possible, thereby reducing data movement and memory copying.

Second, as the system size increases, even the CPU memory may become insufficient, necessitating the use of disk storage backends.
The choice of storage backend is heavily influenced by the number of orbitals and their occupancy ratios.
Given the computational complexity of CCSD, the number of tensors and temporary tensors involved is too large to determine the optimal storage backend in advance and implement the code accordingly.
To address this complexity, we implement a dynamic storage backend selection mechanism alongside a unified code interface.
Tensors in the CCSD procedure are ranked based on their frequency of use and size.
An automatic preprocessing step then determines the appropriate storage location based on the available memory across different backends.
A unified interface is designed to manage tensors across various storage backends, ensuring a consistent interface for tensor computations regardless of the storage backend.
To explain in more detail, tensor contractions involving both CPU and GPU storage backends are handled by the NVIDIA cuTENSORMg library.
For tensor contractions involving disk storage and other tensor operations, our interface performs implicit data movement as required.
This unified interface significantly simplifies the development of the package by abstracting away the complexities of data management, enabling developers to focus on algorithm implementation.

As a result, in ByteQC, the tractable system size for CCSD calculations is the largest within our knowledge to date, while maintaining considerably high efficiency, making it highly suitable for large-scale quantum chemical simulations.

\subsubsection{CCSD(T)}

The energy correlation calculation in CCSD(T) involves contractions between rank-6 tensors, which are too large to store even for modest system sizes.
A common solution is looping over all three virtual orbitals and summing up the energy correlation calculated from each iteration.
In each loop calculation, only rank-3 tensors with sizes proportional to the cube of the number of occupied orbitals are required.
However, in most cases, the contraction of rank-3 tensors only partially occupies the computational resource of the entire GPU\@.
Specifically, according to our benchmark results, this implementation falls into launch-bounded regions, where too many small kernels are launched, causing the overhead of kernel launching (~$\sim$10 $\mu$s per kernel launch) to dominate the computation.

To address this, we utilize customized kernel functions to batch the data required for multiple loops.
The batched data, which fully utilizes the available GPU memory, is then contracted together.
This approach is effectively equivalent to adding an additional broadcasting axis to all tensors involved in the contractions.

The calculation flow of a single iteration for a specific virtual orbital combination $abc$ has also been optimized from the original version~\cite{Scuseria1991} as follows:
\begin{align}
   & \begin{aligned}
       \mathllap{w_{ijk}} & =  P \sum_{d} {[}(ia|bd) t_{jk}^{cd} + (ia|jl) t_{kl}^{bc}{]}
     \end{aligned}
  \\& \begin{aligned}
    \mathllap{v_{ijk}} & = \frac{4*w_{kij} + w_{jki} -2*(w_{kji} + w_{ikj} + w_{jik})}{\alpha(\varepsilon_i+\varepsilon_j+\varepsilon_k - \varepsilon_a-\varepsilon_b-\varepsilon_c)}\label{eq:ccsdt-v}
  \end{aligned}
  \\& \begin{aligned}
    \mathllap{E^{\text{CCSD(T)}}_{abc}} & = \sum_{ijk} w_{ijk} v_{ijk}
  \end{aligned}
  \\& \begin{aligned}
    \mathllap{x_k} & = 0.5 P \sum_{ij} t_{ij}^{ab} v_{ijk}\label{eq:ccsdt-x}
  \end{aligned}
  \\& \begin{aligned}
    \mathllap{y_k} & = 0.5 P \sum_{ij} (ia|jb) v_{ijk}\label{eq:ccsdt-y}
  \end{aligned}
  \\& \begin{aligned}
    \mathllap{E^{\text{CCSD(T)}}_{abc}} & = E^{\text{CCSD(T)}}_{abc} + \sum x_k F_{ck} +\sum y_k t_{k}^c,\label{eq:ccsdt-e}
  \end{aligned}
\end{align}
where $t$ is the single or double excitation amplitude in CCSD, $\alpha$ is a constant to remove double-counting, $\varepsilon$ is the molecular energies, $F$ is the Fock matrix, and $P$ means the sum over all permutations of $abc$.
By this way, only extra memory as the size of $w$ and $v$ is needed in total for the iterations.
No additional memory is required as contractions are performed using a transpose-free algorithm and the denominator in Eq.~(\ref{eq:ccsdt-v}) is computed in-place by a custom GPU kernel.
The rank-1 tensors $x$ and $y$ are much smaller than the rank-3 tensor $v$ and can reuse the memory of $v$.
Additionally, in Eqs.~(\ref{eq:ccsdt-x},\ref{eq:ccsdt-e}), the energy is computed by contracting $t_{ij}^{ab}$, $v$, and $F$ together.
While contracting $t_{ij}^{ab}$ with $F$ first also yields the correct energy, it is inefficient in terms of both memory usage and computational cost.
Similarly, contracting $(ia|jB)$, $v$, and $t_k^c$ in Eqs.~(\ref{eq:ccsdt-y},\ref{eq:ccsdt-e}) in the wrong order leads to the same inefficiencies.

\subsection{Systematically improvable embedding}

Quantum embedding~\cite{Sun2016}, as a fragmentation-based method, offers a potential solution to solve large-scale systems at higher accuracy.
The large system is first fragmented according to low-level solutions (typically at mean-field level).
The divided subsystems are then solved using a high-level solver (post-HF methods such as RPA, CCSD, or CCSD(T)), and the results from different subsystems are collected and recombined to construct the solution for the entire system.
In this way, the range of system sizes that can be handled computationally is significantly expanded, driving the rapid development of quantum embedding techniques over the past few years~\cite{Nusspickel2022,Knizia2012,Ye2024,Lau2021,Ye2019}.
However, the issue of the lack of systematic improvability remains unresolved in the original quantum embedding method,
which stems that it is hard to obtain an exact description of entanglement between fragments and their environment at the mean-field level.

\begin{figure}[htb!]
  \centering
  \inputtikz{sie}{}{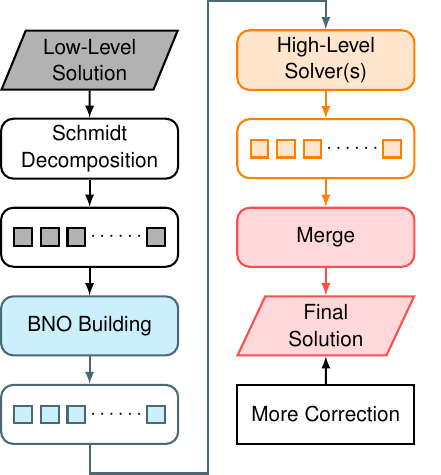}
  % \vspace{8mm}
  \caption{The general framework of SIE.\label{fig:SIE_framework}}
\end{figure}

In this work, we focus on a variant of DMET, the SIE method, which is introduced in 2022~\cite{Nusspickel2022} and proposed to address this problem.
The general framework of SIE is illustrated in Fig.~\ref{fig:SIE_framework}.
Similarly to DMET, SIE requires a low-level solution for the entire system combined with a predefined partition strategy.
Then, the Schmidt decomposition is used to break down the whole system into subsystems following the common practice of obtaining \textit{fragment} and \textit{bath}~\cite{Wouters2016}.
Based on the fragment, the SIE further constructs the \textit{cluster} by adding a special type of orbitals to the subspace, named as bath natural orbitals~(BNOs), which are selected by their MP2 interactions with their corresponding subspaces to utilize the information beyond mean-field level.
In our test example, the BNO select threshold is set as $10^{-8}$ to ensure that the accuracy is high enough with a reasonable computational cost.
Note that the smaller the threshold, the larger the cluster becomes and, correspondingly, the better accuracy is achieved.
RPA is also proposed as a method to construct BNOs~\cite{Scott2024}.
After constructing the cluster, it is typically solved using correlated methods, with MP2 and CCSD being common options.
Finally, solutions from all clusters are merged to form the solution for the entire system.
The merging methods are discussed in detail in the literature~\cite{Nusspickel2023}, and the partition wave function density matrix approach is highly recommended because it introduces unique cluster interactions and naturally ensures the N-representability of the solution~\cite{Nusspickel2023}.
To obtain better accuracy, further corrections can also be considered.
For instance, there are correlations outside the cluster due to the truncation of BNOs, which introduce a so-called bath truncation error.
This error can be corrected through an additional MP2 calculation of the entire system.

Due to the limitation of the merging method, previous SIE calculations were performed with CCSD as the most accurate high-level solver.
After several efforts and attempts, a modified CCSD(T), tailored for SIE, is adopted, ultimately forming a composite method referred to as SIE+CCSD(T), which pushes the accuracy of SIE calculations to the golden-standard level.
A detailed description of the SIE+CCSD(T) method and additional corrections can be found in our previous work~\cite{Huang2024}.
However, the $O(N^7)$ computational complexity of CCSD(T) makes its application prohibitively expensive, even when used as a high-level solver.
In this work, leveraging the efficient GPU implementation of the common algorithms, including this modified CCSD(T), we can further efficiently improve the accuracy of the SIE method to the highest CCSD(T) level.

To implement SIE, all the individual components provided by our package are required as shown in Fig.~\ref{fig:pkg}, making it an ideal test case to showcase the capabilities of our package.
Due to the fragmentation-based nature of SIE, the computations for different clusters are independent, making it inherently well-suited for parallel processing across multiple nodes.
In ByteQC, SIE framework is realized in a subpackage named ``embyte'', and is implemented using the message passing interface with all low-level and high-level solvers GPU-accelerated.
Currently, ``embyte'' fully supports the calculation of SIE+CCSD(T), including the construction of BNOs with MP2 methods and the merge of cluster solutions with the partition wave function density matrix approach.

\section{\MakeUppercase{Benchmark result}\label{sec:results}}

The GPU benchmarks are run with an NVIDIA A100 GPU with 80 GB of global memory.
Multiple GPUs are interconnected through NVLink, enabling high-speed data transfer and enhanced communication bandwidth.
The versions of related software utilized in the benchmarks are listed below.

\begin{itemize}
  \item CUDA 12.6
  \item CuPy (main branch in GitHub to support cuTENSORMg)
  \item NCCL 2.23.4
  \item cuTENSOR 2.1.0
  \item PySCF 2.5.0
\end{itemize}

There is no universal method for fairly comparing the efficiency of GPU and CPU codes.
Two commonly adopted metrics, the speedup of wall time and the equivalent number of CPU cores corresponding to a GPU, are strongly influenced by hardware details and the number of CPU cores used in the test.
For this benchmark, we chose 100 logical cores of an Intel\textsuperscript{\textregistered} Xeon\textsuperscript{\textregistered} Platinum 8336C CPU to benchmark our results, as 100 cores are in the same order of magnitude as the typical size of a single node, and provide a convenient basis for translating between speedup and equivalent CPU cores.

We choose the Benzene crystal with various basis sets for the speedup benchmark of ``cupbc'', while a catalyst system comprising up to 100\,000 orbitals is used to evaluate its multi-GPU scaling performance.
Water clusters with varying numbers of water molecules are selected for the benchmarks of ``cuobc'', ``cump2'', and ``cucc''.
We increase the number of water molecules until hardware resource limitations or computational time exceeding 10\,000 seconds, which would correspond to more than one day for the entire calculation, assuming that 10 iterations are required for convergence.
To demonstrate the SIE, we calculated the adsorption energy of water clusters and the interaction energy between a water monomer and a hexagonal boron nitride~(h-BN) surface.

\subsection{Benchmark of subpackage ``cupbc''}

As illustrated in Fig.~\ref{fig:cupbc-s}, for periodic benzene crystals, our codes show about 10$\sim$20 times speedup compared to the CPU implementations with 100 cores in different basis sets.
The speedup for high angular momentum is slightly lower than that for low angular momentum, as more complex recursion is required, which is less efficient on the GPU\@.
Furthermore, to show the capacity of our codes, we benchmark our codes in a very large catalyst system up to 100\,000 orbitals (4$\times$4$\times$5 k-mesh with 1\,250 basis), a carbon monoxide on the copper surface, in Fig.~\ref{fig:cupbc-m}.
The Fock matrix of such huge systems can be done with one A100 GPU within 43.2 hours.
Compared to the out-core version, the full and momentum-conserved CD-ERIs of this system are as large as 2\,760.9 TB and 34.5 TB, respectively.
Building the Fock matrix requires reading the CD-ERIs twice, which takes approximately 3\,141.3 hours for full CD-ERIs and 39.3 hours for momentum-conserved CD-ERIs, given a typical high-speed hard disk with a read speed of 500 MB/s.
This estimate even excludes the computational time and complexity of processing such large data under memory constraints.
The ideal linear scaling can also be achieved as shown in Fig.~\ref{fig:cupbc-m} which indicates such calculation can be further reduced to 5.5 hours with 8 A100 GPUs.
Therefore, the on-the-fly calculation on GPU is concluded to be the most suitable scheme for large-scale periodic system mean-field research.

\begin{figure}[htb!]
  \centering
  \inputtikz{cupbc}{s, m}{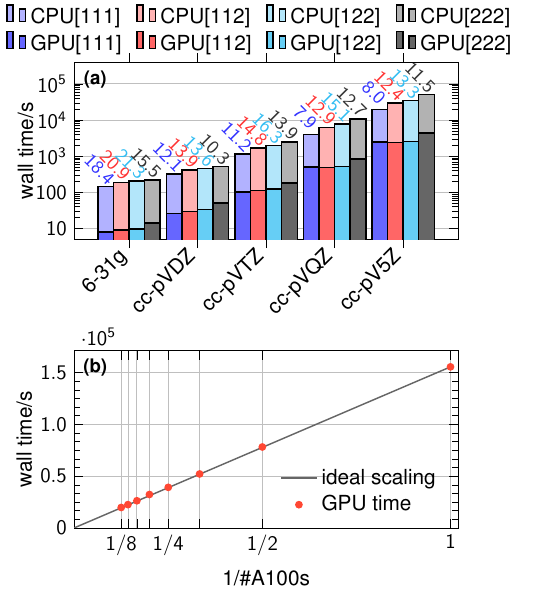}
  \caption{\textbf{(a)} Benchmarks of Fock matrices building time on a Benzene crystal with k-mesh equal to $1\times1\times1$, $1\times1\times2$, $1\times2\times2$, and $2\times2\times2$.
    The total numbers of basis are 264, 456, 1\,056, 2\,040, and 3\,504 respectively.
    The A100 GPU and a 100-core CPU is used.
    The numbers in the figures indicate the speedup of GPU implementation.
    \textbf{(b)} The wall time of the Fock matrix building for system with a carbon monoxide on the copper surface for different number of GPUs.
    The k-mesh is $4\times4\times5$ and the number of basis is 1\,250, resulting total 100\,000 orbitals.
    The black line indicates the ideal linear scaling.\label{fig:cupbc}}
\end{figure}

\subsection{Benchmark of subpackage ``cuobc''}

\begin{figure}[H]
  \inputtikz{bench}{cuobc-s, cuobc-t, cuobc-m, cump2-s, cump2-t, cump2-m, ccsd-s, ccsd-t, ccsd-m, ccsdt-s, ccsdt-t, ccsdt-m}{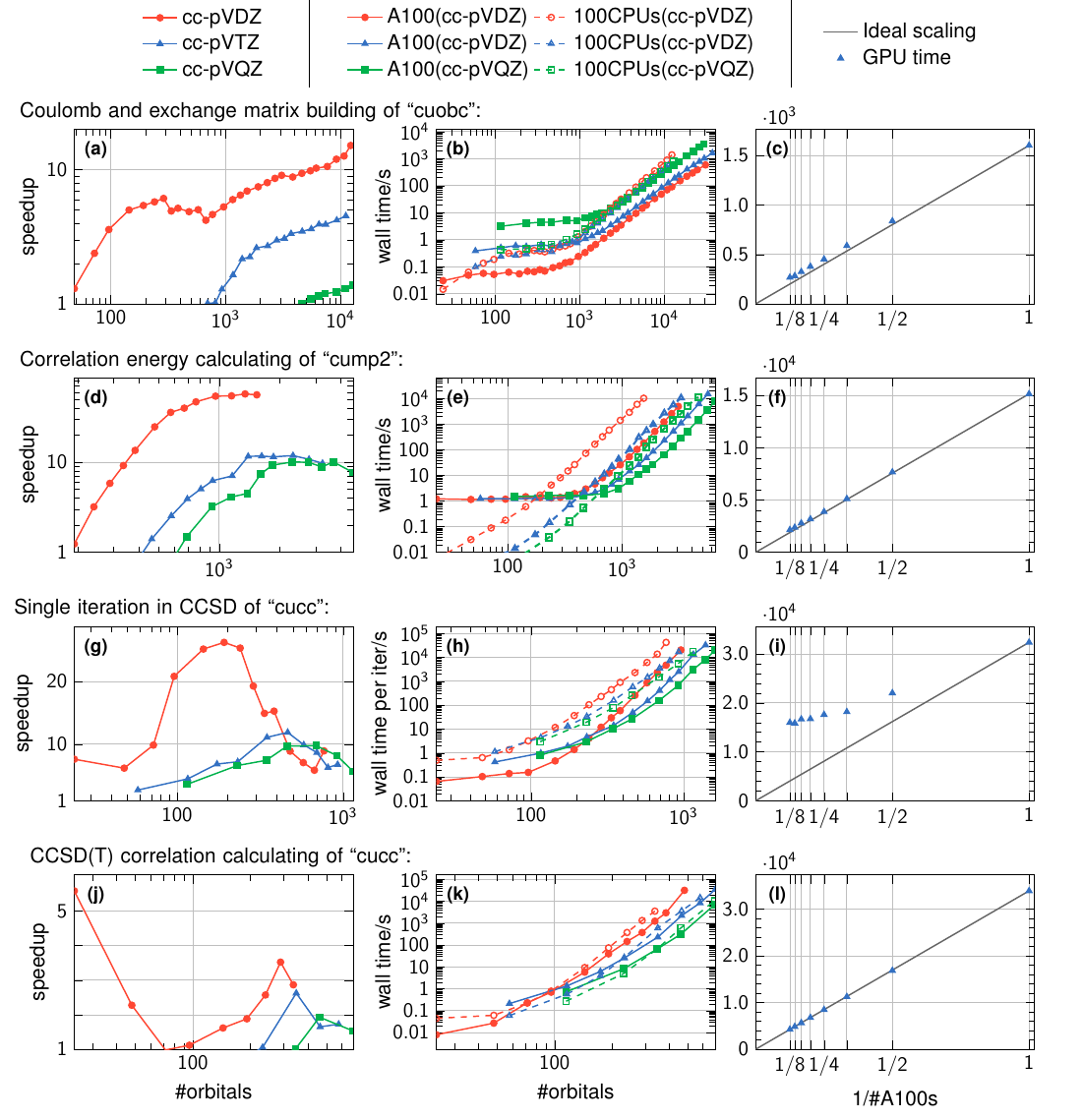}
  \vspace{-0.5cm}
  \caption{Benchmarks on water clusters with different basis sets.
    \textbf{(a,d,g,j)} The speedups of an A100 GPU compare to a 100-core CPU\@.
    \textbf{(b,e,h,k)} The wall times of CPU and an GPU for different number of orbitals.
    \textbf{(c,f,i,l)} The wall times with different GPU numbers for the cc-pVTZ basis set.
    The orbital numbers for multi-GPU tests are 37\,120, 5\,568, 1\,392, and 696 respectively.
    The black line indicates the ideal scaling.\label{fig:bench}}
\end{figure}

We benchmark OBC mean-field calculations by building Coulomb and exchange matrices with different basis sets as illustrated in Figs.~\ref{fig:bench}\subref{fig:bench-cuobc-s}-\subref{fig:bench-cuobc-m}.

As shown in Fig.~\ref{fig:bench-cuobc-s}, the maximum speedups are about 15.2, 4.5, and 1.4 for the basis sets cc-pVDZ, cc-pVTZ, and cc-pVQZ, respectively.
The acceleration drop for larger basis sets could be related to two reasons.
One reason is that for larger basis sets, high angular momentum will involve more complex recursion in the ERI calculation, which is inefficient when compared to CPU\@.
Another reason is related to the different prescreen efficiency for CPU and GPU\@.
In the CPU version, the threads are independent, allowing for much flexibility in prescreening.
In contrast, the threads in the GPU version are grouped together, and prescreening is applied either for all threads simultaneously or not at all.
If prescreening is turned off for both the CPU and GPU codes, the speedup is found to be more than an order of magnitude larger compared to the case with prescreening.

The same scaling is observed for the wall times of both CPU and GPU calculations with different numbers of orbitals in Fig.~\ref{fig:bench-cuobc-t} while the wall times of GPUs are much smaller compared to that of 100-core CPUs.
For our GPU codes, the maximum number of orbitals that can be handled is up to 37\,120.
Beyond this point, the Coulomb and exchange matrices occupy a significant portion of the total 80 GB GPU memory, making subsequent computations infeasible.

We also demonstrate that nearly ideal linear scaling with respect to multiple GPUs is achieved, as shown in Fig.~\ref{fig:bench-cuobc-m}, which further accelerates quantum chemistry simulations based on mean-field calculations.
The slight deviation from ideal scaling is attributed to the imbalance in workload distribution across different GPUs.

\subsection{Benchmark of subpackage ``cump2''}

The MP2 benchmark results are illustrated in Figs.~\ref{fig:bench}\subref{fig:bench-cump2-s}-\subref{fig:bench-cump2-m}.
As illustrated in Fig.~\ref{fig:bench-cump2-s}, we achieve a maximum speedup of 56.0 times for the cc-pVDZ basis sets, and approximately 10 times for the larger basis sets~(cc-pVTZ, cc-pVQZ) compared to a 100-core CPU\@.
In Fig.~\ref{fig:bench-cump2-t}, plateaus of wall time are observed when system size is smaller than several hundreds.
That is attributed to the overhead of multi-processing, and is negligibly small ($<1$ s).
The differences between different basis sets are caused by the different ratio between the occupied and virtual orbitals.
The largest fully benchmarked system involves 6\,440 orbitals, while a system with 11\,040 orbitals has been tested and confirmed to be manageable in memory, albeit with a longer runtime.
Beyond this, GPU memory becomes insufficient.
The ideal linear scaling is also achieved as shown in Fig.~\ref{fig:bench-cump2-m}.

\subsection{Benchmark of subpackage ``cucc''}

\subsubsection{Benchmark of CCSD}

For CCSD, we set the same CPU memory limit of 240 GB for both GPU version in ByteQC and the CPU version in PySCF to ensure a fair comparison.
As shown in Fig.~\ref{fig:bench-ccsd-s}, our CCSD implementation achieves a speedup of up to 20$\sim$30 times compared to a 100-core CPU\@.
It should be pointed out that, in order to better balance larger systems with faster computations without introducing complex conditional branches, certain trade-offs are made.
The speedups can certainly be further improved where all tensors fit entirely within GPU memory.

It is well known that the occupied ratio greatly affects the performance of CCSD and the largest system that it can handle.
As shown in Fig.~\ref{fig:bench-ccsd-t}, the larger basis set (smaller occupied ratio) is cheaper for CCSD given the same total orbitals.
In our test, the large systems being benchmarked are as large as 960, 1\,392 and 1\,610 for cc-pVDZ, cc-pVTZ, and cc-pVQZ with occupied orbital ratios 20.8\%, 8.6\%, and 4.3\% respectively.
The 1\,610 orbitals is the largest system reported in the literature to the best of our knowledge.
For even larger systems, some tensors scaling as the third power of the number of orbitals may not fit into GPU memory, which is beyond the scope of our current interests.

CCSD calculation is inherently serial and difficult to divide into independent parts.
In our package, only the contraction between tensors will benefit from adding more GPUs through the cuTENSORMg library.
The data movement between different backends also prevents achieving ideal scaling as shown in Fig.~\ref{fig:bench-ccsd-m}. However, a speedup of approximately 1.5 times is achieved when adding another GPU, and the speedup gradually increases to 2.0 as the number of GPUs continues to grow.

\subsubsection{Benchmark of CCSD(T)}

As shown in Fig.~\ref{fig:bench-ccsdt-s}, the maximum speedup of our CCSD(T) implementation over the CPU one in PySCF is 5.9, 1.7, and 1.5 for basis sets cc-pVDZ, cc-pVTZ, cc-pVQZ, respectively.
The relatively lower speedup is due to the heavy memory read overhead, and a more efficient version is currently under development.
In Fig.~\ref{fig:bench-ccsdt-t}, the scaling of CCSD(T) varies with different basis sets for both CPU and GPU codes, which is attributed to implementation details.
In our implementation, the available GPU memory determines the choice of batch size, which subsequently determines the number of iterations.
Larger orbitals result in more iterations with smaller batch sizes, leading to less computation per iteration and lower GPU utilization.
A system with up to 696 orbitals has been fully benchmarked, while our package has been tested to be capable of handling up to 1\,380 orbitals based on tests of the first few iterations.
The ideal linear scaling is demonstrated to be achievable, as shown in Fig.~\ref{fig:bench-ccsdt-m}, indicating that large systems can be completed in a reasonable amount of time with the use of additional GPUs.

\subsection{Benchmark of subpackage ``embyte''}

Unlike the standalone subpackages, where the CPU and GPU versions follow the same computational flow and validation has been confirmed by ensuring identical results within numerical error, no equivalent implementation exists for SIE+CCSD(T).
Therefore, the validation of the ``embyte'' subpackage is carefully benchmarked in this section.

\begin{figure}[htb!]
  \centering
  \inputtikz{wc}{}{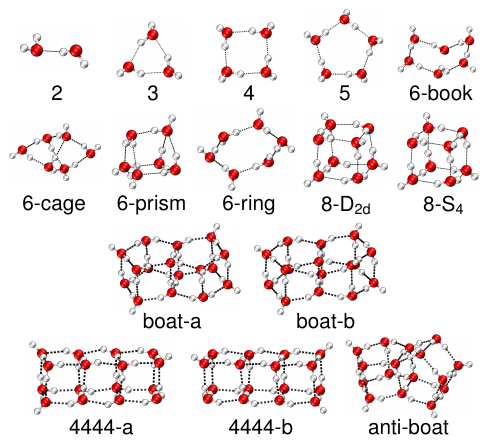}
  \caption{The structures for water clusters (H$_2$O)$_{\text{2-6,8,16}}$.\label{fig:wc_structures}}
\end{figure}

Water clusters, denoted as (H$2$O)${_m}$, where $m$ represents the number of water molecules, are considered ideal for this purpose.
Accurate modeling of their inherently long-range interactions introduced by the extensive hydrogen-bonding network~\cite{Yuan2017,Apra2009,Alfe2013,Yoo2010,Howard2013,Santra2011,Gillan2013,Miliordos2015,Wang2014} necessitates methods achieving at least CCSD(T)-level precision, as established in previous studies~\cite{Howard2013, Yoo2010, Gillan2013, Miliordos2015}.
Our benchmark set includes water clusters (H$_2$O)$_{m}$ with $m$ = 2-6, 8, and 16, with representative structures illustrated in Fig.~\ref{fig:wc_structures}.
Beyond validation, by leveraging the advantages of SIE, namely high accuracy with relatively low computational cost, we also generate higher-quality data using larger basis sets to benefit the scientific community.

By default, SIE+CCSD(T) uses the partitioned wavefunction reduced density matrix~\cite{Nusspickel2023} to calculate the energy, with the BNO truncation threshold set as $10^{-8}$.
MP2 is used to address the bath truncation error~\cite{Huang2024}.
Each water monomer is treated as an individual fragment.

\subsubsection{Benchmark on (H$_2$O)$_{\text{2-6,8}}$}

For the water clusters (H$_2$O)$_{\text{2-6,8}}$, the binding energies $D_{m}$ are defined as~\cite{Miliordos2015}
\begin{equation}
  D_m = E_{{(\text{H}_{2}\text{O})}_{m}} - mE_{\text{H}_{2}\text{O}},
\end{equation}
where $E_{{(\text{H}_{2}\text{O})}_{m}}$ or $E_{\text{H}_{2}\text{O}}$ denotes the total energy for (H$_2$O)$_{m}$ or a single water monomer.

All calculations in the above formula are expected to be extrapolated to the complete basis set~(CBS) limit, which is obtained using two extrapolation strategies~\cite{Miliordos2015}.

The first strategy extrapolates the results from multiple Gaussian basis sets to the CBS limit, referred to as multipoint extrapolation.
Specifically, four progressively larger basis sets, aug-cc-pV(D,T,Q,5)Z~\cite{Dunning1989}, are used.
This method is applied to deal with (H$_2$O)$_{\text{2-4}}$ water clusters.
For larger (H$_2$O)$_{\text{5,6,8}}$ water clusters, the computation cost becomes too expensive to obtain the results of canonical CCSD(T) at basis sets larger than the aug-cc-pVTZ\@.
Consequently, the second strategy, the MP2-assisted scaling method, is employed as a compromise solution.
The results at the CBS limit are obtained by scaling the CCSD(T) results with the aug-cc-pVTZ basis set by a ratio, which is determined by taking the ratio of MP2 results at the CBS limit to those with the aug-cc-pVTZ basis set and then averaging across the first 10 water clusters (H$2$O)$_{\text{2-6,8}}$ in Fig.~\ref{fig:wc_structures}:
\begin{equation}
  \label{eq:MP2_assist_CBS_extrap}
  D_m^\text{CBS,CCSD(T)} = D_m^{\text{aug-cc-pVTZ,CCSD(T)}} \times \frac{1}{k} \sum_{\text{all clusters}} \frac{D_n^{\text{CBS,MP2}}}{D_n^{\text{aug-cc-pVTZ,MP2}}},
\end{equation}
where $n$ represents the number of water molecules in each water cluster, and $k$ denotes the number of water clusters in the dataset with $k=10$ in this study.
However, MP2 exhibits size inconsistency making direct extrapolation across different systems problematic and less reliable than multipoint extrapolation.

In our benchmark, we first adopt two similar strategies for SIE+CCSD(T), using only the aug-cc-pV(T,Q)Z basis sets for the multipoint extrapolation in MP2/SIE+CCSD(T) calculations.
More specifically, the extrapolation~\cite{Halkier1999, Halkier1998} formulas for the HF energy and the correlation energy are
\begin{align}
  E^\text{CBS,HF}   & = E^\text{x,HF} - \frac{E^\text{x,HF}-E^\text{y,HF}}{1-e^{-1.637}},\label{eq:HF_two_point_extrap}  \\
  E^\text{CBS,corr} & = \frac{x^3 E^\text{x,corr} - {y}^3 E^\text{y,corr}}{x^3 - {y}^3},\label{eq:corr_two_point_extrap}
\end{align}
with $x$ and $y$ the $\zeta$-cardinality of the basis sets.
As shown in Table~\ref{tab:wc_2_8}, the maximum difference between the results of SIE+CCSD(T) and the results of CCSD(T) in Ref.~\cite{Yoo2010} is merely 0.14~kcal/mol for water clusters (H$_2$O)$_{\text{2-4}}$ and 0.5~kcal/mol for (H$_2$O)$_{\text{5,6,8}}$.
These differences fall within chemical accuracy, demonstrating the high reliability and precision of SIE+CCSD(T).

\begin{table}[htb!]
  \centering
  \begin{threeparttable}
    \caption{
      The binding energies (kcal/mol) in CBS limit for (H$_2$O)$_{\text{2-6,8}}$.\label{tab:wc_2_8}
    }
    \begin{tabular}{cccccc}
      \toprule\midrule
      Structure         & Ref.~\cite{Miliordos2015} CCSD(T)$^a$ & SIE+CCSD(T)$^a$ & SIE+CCSD(T)$^b$ \\
      \midrule
      2                 & -4.95                                 & -4.89           & -4.89           \\
      3                 & -15.68                                & -15.59          & -15.59          \\
      4                 & -27.25                                & -27.11          & -27.11          \\
      5                 & -35.9$\pm$0.4                         & -35.7$\pm$0.5   & -35.67          \\
      6-book            & -45.6$\pm$0.2                         & -45.5$\pm$0.7   & -44.82          \\
      6-cage            & -46.0$\pm$0.5                         & -45.5$\pm$0.7   & -45.17          \\
      6-prism           & -46.3$\pm$0.5                         & -46.3$\pm$0.7   & -45.52          \\
      6-ring            & -44.2$\pm$0.4                         & -43.9$\pm$0.6   & -44.14          \\
      8-D$_{\text{2d}}$ & -73.3$\pm$0.7                         & -73.7$\pm$1.1   & -71.40          \\
      8-S$_4$           & -73.2$\pm$0.7                         & -73.6$\pm$1.1   & -71.37          \\
      \midrule\bottomrule
    \end{tabular}
    \begin{tablenotes}
      \small
      \item[$a$] Using multipoint extrapolation for (H$_2$O)$_{\text{2-4}}$ and MP2-assisted scaling method for (H$_2$O)$_{\text{5,6,8}}$.
      \item[$b$] Using multipoint extrapolation with aug-cc-pV(T,Q)Z basis set.
    \end{tablenotes}
  \end{threeparttable}
\end{table}

Furthermore, although larger basis sets are computationally prohibitive for canonical CCSD(T) in (H$_2$O)$_{\text{5,6,8}}$, they are easily manageable with SIE+CCSD(T).
The results obtained using the multipoint extrapolation method with aug-cc-pV(T,Q)Z basis sets to reach the CBS limit are also calculated and listed in Table~\ref{tab:wc_2_8}.
The discrepancy between the results from multipoint extrapolation and those from the MP2-assisted scaling method with SIE+CCSD(T) reaches up to 2.3~kcal/mol, highlighting the MP2 inconsistency problem and demonstrating the superiority of the SIE+CCSD(T) method.

\subsubsection{Benchmark on (H$_2$O)$_{\text{16}}$}

For the water cluster (H$_2$O)$_{16}$, five isomer structures presented in the last two rows of Fig.~\ref{fig:wc_structures} have been selected.
These five isomers are nearly degenerate, with the energy differences not exceeding 1 kcal/mol, posing a challenge for the method's accuracy in distinguishing their relative energies.
The results of the reference CCSD(T) and SIE+CCSD(T) calculations using the aug-cc-pVTZ basis set are shown in Table~\ref{tab:wc_16_results}, with all energies referenced to the lowest value among the five isomers, i.e., the energy of the 4444-a isomer.
The largest difference between SIE+CCSD(T) and the reference is within 0.19 kcal/mol, which can be considered an excellent agreement.

Owing to high computational costs, the reference canonical CCSD(T) lacks results for the aug-cc-pVQZ basis set, preventing multipoint extrapolation to the CBS limit.
However, our GPU-accelerated SIE+CCSD(T) method overcomes this limitation.
The relative energies of all five isomers are further calculated using aug-cc-pVQZ and extrapolated to the CBS limit, which are presented in Table~\ref{tab:wc_16_results}.
The total SIE+CCSD(T) time consumption is estimated using a single A100 GPU for all water clusters, and listed in Table~\ref{tab:time_consumption}.

\begin{table}[htb!]
  \centering
  \begin{threeparttable}
    \caption{Relative energies (kcal/mol) for five (H$2$O)$_{16}$ isomers, all aligned to the minimum energy.\label{tab:wc_16_results}
    }
    \begin{tabular}{ccccccc}
      \toprule\midrule
      \multirow{2}{*}{Structure} & Ref.~\cite{Yoo2010} CCSD(T)     & SIE+CCSD(T) & SIE+CCSD(T)$^a$ \\
      \cline{2-3}                & \multicolumn{2}{c}{aug-cc-pVTZ} & CBS                           \\
      \midrule
      boat-a                     & 0.255                           & 0.232       & 0.474           \\
      boat-b                     & 0.421                           & 0.409       & 0.641           \\
      4444-a                     & 0                               & 0           & 0               \\
      4444-b                     & 0.542                           & 0.350       & 0.538           \\
      anti-boat                  & 0.511                           & 0.438       & 0.915           \\
      \midrule\bottomrule
    \end{tabular}
    \begin{tablenotes}
      \small
      \item[$a$] Using multipoint extrapolation to CBS limit.
    \end{tablenotes}
  \end{threeparttable}
\end{table}

\begin{table}[htb!]
  \centering
  \caption{
    The time consumption of SIE+CCSD(T) and related information for different systems.
    The basis set is aug-cc-pVQZ for water clusters and cc-pVQZ for H$_2$O@h-BN.
    The time consumption is estimated on single A100 GPU.\label{tab:time_consumption}
  }
  \begin{tabular}{cccc}
    \toprule\midrule
    System                           & System size & Average cluster size & Time/hour \\
    \midrule
    (H$_2$O)$_1$                     & 172         & 169.0                & 0.1       \\
    (H$_2$O)$_2$                     & 344         & 329.0                & 1.0       \\
    (H$_2$O)$_3$                     & 516         & 453.0                & 7.0       \\
    (H$_2$O)$_4$                     & 688         & 464.0                & 9.5       \\
    (H$_2$O)$_5$                     & 860         & 471.8                & 12.7      \\
    (H$_2$O)$_6$, 6-book             & 1032        & 501.8                & 21.7      \\
    (H$_2$O)$_6$, 6-cage             & 1032        & 527.5                & 29.8      \\
    (H$_2$O)$_6$, 6-prism            & 1032        & 537.0                & 33.3      \\
    (H$_2$O)$_6$, 6-ring             & 1032        & 479.0                & 16.3      \\
    (H$_2$O)$_8$, 8-D$_{\text{2d}}$  & 1376        & 548.0                & 38.3      \\
    (H$_2$O)$_8$, 8-S$_4$            & 1376        & 548.0                & 39.1      \\
    (H$_2$O)$_{16}$, boat-a          & 2752        & 567.9                & 199.3     \\
    (H$_2$O)$_{16}$, boat-b          & 2752        & 568.1                & 217.8     \\
    (H$_2$O)$_{16}$, 4444-a          & 2752        & 573.6                & 242.7     \\
    (H$_2$O)$_{16}$, 4444-b          & 2752        & 574.0                & 263.0     \\
    (H$_2$O)$_{16}$, anti-boat       & 2752        & 571.5                & 240.0     \\
    H$_2$O@h-BN, $6\times6\times1$   & 3926        & 460.0                & 795.1     \\
    (H$_2$O)@h-BN, $6\times6\times1$ & 3926        & 455.5                & 725.0     \\
    H$_2$O@(h-BN), $6\times6\times1$ & 3926        & 135.0                & 0.3       \\
    \midrule\bottomrule
  \end{tabular}
\end{table}

\section{Application\label{sec:app}}

Monomer adsorption on surfaces is crucial to numerous chemical processes~\cite{Schedin2007,Farha2010,Secchi2016,Siria2017}, yet most adsorption predominantly relies on van der Waals interactions resulting in weak and delocalized binding.
As interacting energy constitutes a dominant component of adsorption energy, its accurate determination requires high-accuracy calculations for extremely large models.
Such a demand aligns well with the strengths of SIE+CCSD(T) in delivering sufficient accuracy with high efficiency.
Therefore, interacting energy calculations are demonstrated as a practical application in this work.
Building upon our previous work~\cite{Huang2024}, which established the potential of SIE+CCSD(T) for accurate interacting energy calculations, the present study introduces a new demonstrative case to further validate its capabilities: the interacting energy $E_{\text{int}}$ of a water monomer adsorbed on h-BN\@ surface (Fig.~\ref{fig:H2O_BN-struct}), denoted as H$_2$O@h-BN\@.
It has also been extensively studied using other high-accuracy quantum chemical methods~\cite{AlHamdani2017,Gruber2018,Lau2021}.

\begin{figure}[htb!]
  \centering
  \inputtikz{H2O_BN}{struct,bulk}{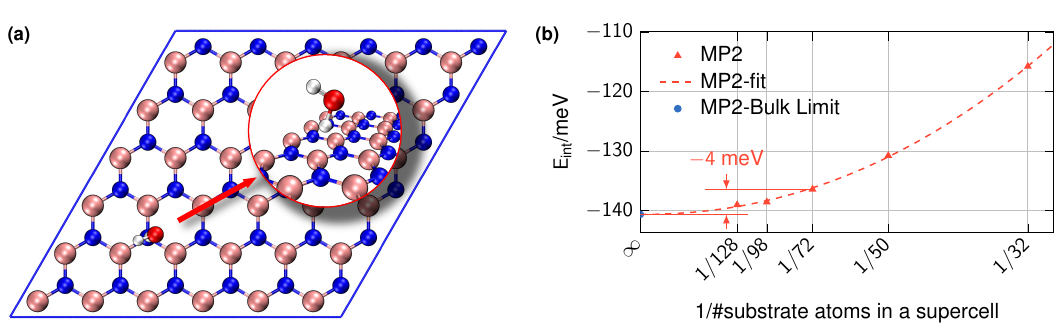}
  \caption{\textbf{(a)} The structure for H$_2$O@h-BN\@, using $6 \times 6 \times 1$ supercell.
    The structure is taken from Ref.~\cite{AlHamdani2017}.
    \textbf{(b)} The interacting energies $E_{\text{int}}$ calculated by MP2 as the increasing total numbers of atoms in supercell of substrates.
    The Infinite symbol represents the bulk limit.\label{fig:H2O_BN}}
\end{figure}

The definition of interacting energy $E_{\text{int}}$ with the common counterpoise correction~\cite{Boys1970} for basis set superposition error reads
\begin{align}
  E_{\text{int}} = E_{\text{H}_2\text{O@h-BN}} - E_{\text{H}_2\text{O@(h-BN)}} - E_{(\text{H}_2\text{O})\text{@h-BN}}
\end{align}
with $E_{\text{H}_2\text{O@h-BN}}$ representing the total energy for H$_2$O@h-BN, and $E_{\text{H}_2\text{O@(h-BN)}}$ representing the total energy calculated using only the h-BN\@ basis, excluding its electrons and nuclear cores, and similarly for $E_{(\text{H}_2\text{O})\text{@h-BN}}$.
The interacting energy between a water monomer and h-BN\@ is calculated under periodic boundary conditions using the cc-pV(T,Q)Z basis set with multipoint CBS extrapolation.
The SIE settings follow those used in the benchmark section for water cluster calculations.
The water monomer and each individual B or N atom in h-BN are treated as separate fragments.

A convergence test of the interacting energy with increasing substrate size is conducted at MP2 level (see Fig.~\ref{fig:H2O_BN-bulk}), and the method for bulk limit extrapolation in Ref.~\cite{Huang2024} is adopted.
We observe that when the substrate consists of 72 atoms ($6\times6\times1$ supercell), the difference compared to the extrapolated bulk limit is only -4~meV as shown in \ref{fig:H2O_BN-bulk}, indicating the interacting energy approximately converges at this substrate size.
Therefore, we selected a $6\times6\times1$ h-BN\@ substrate for SIE+CCSD(T) calculations, compensating for the -4~meV MP2-level difference to estimate the bulk-limit interaction energy.
Our MP2 and SIE+CCSD(T) data align perfectly with the reference value within 1~meV from Ref.~\cite{Gruber2018} at the triple-$\zeta$ basis set level, despite the use of more diffuse basis sets in the reference.

Due to computational limitations, the reference k-CCSD(T) could not employ quadruple-$\zeta$ basis sets, while SIE+CCSD(T) can still handle.
With the cc-pV(T,Q)Z basis set extrapolation, the interaction energy of H$_2$O@h-BN at CBS limit is found to be -119~meV at $6\times6\times1$ h-BN supercell.
A more realistic result with both CBS limit and the bulk limit, -123~meV, is then obtained by adding the -4~meV MP2 correction, representing the most concrete CCSD(T)-level result.
The SIE+CCSD(T) consumptions estimated on a single A100 GPU for this part are also listed in Table~\ref{tab:time_consumption}.
In summary, by utilizing SIE+CCSD(T), we have achieved improved accuracy at the CCSD(T) level for large systems through our efficient GPU implementation.
It is expected that the ByteQC package will demonstrate its advantages in further research on large-scale systems and make significant contributions to the quantum chemistry community.

\section*{Conclusion and outlook}

Quantum chemistry simulations require substantial computational resources and diverse algorithms to manage systems of varying sizes.
To address this challenge, we introduce ByteQC, an open-source quantum chemistry package designed to accelerate simulations using GPU techniques, particularly for large-scale systems.

ByteQC implements several widely used algorithms, which have been benchmarked for high efficiency and demonstrated to effectively handle large systems.
For molecular mean-field calculations, we observe a maximum speedup of 15.2 times using an A100 GPU compared to a 100-core CPU, with the largest system benchmarked containing up to 30\,720 orbitals.
Additionally, a periodic catalyst system with 80 kmesh-points and 1\,250 basis functions, resulting in a total of 100\,000 orbitals, is within the capabilities of our package.
For such periodic mean-field calculations, speedups of 10 to 20 times are achieved under various scenarios.
In terms of MP2 correlation, our GPU-accelerated code delivers up to 56.0 times the speedup compared to the CPU version, with the largest system confirmed to be manageable containing 11\,040 orbitals.
Efficient memory management allows the largest CCSD calculation in our benchmarks to involve 1\,610 orbitals, achieving a maximum speedup of 20.8 times during all benchmarks.
A speedup of up to 5.9 times is observed for CCSD(T) calculations, with systems containing up to 1,380 orbitals demonstrated to be manageable through the execution of the first few iterations.
All functionalities support multi-GPU setups and exhibit good scalability with the number of GPUs used.

An advanced feature, the SIE framework, is also provided in ByteQC, which integrates all the aforementioned components.
The accuracy of SIE+CCSD(T) is rigorously benchmarked in water clusters, achieving sub-chemical accuracy agreement with canonical CCSD(T) results.
Subsequently, to illustrate its practical potential, we employed H$_2$O@h-BN\@ as a representative application, presenting the complete workflow for interaction energy calculations.
Our results achieved meV-level consistency with previous works.
Leveraging the computational efficiency of the SIE and the robust capabilities of ByteQC, we further performed more accurate calculations using extended basis sets, which demonstrate that our package is capable of giving more accurate results beyond the scope of existing research.

ByteQC also includes a variety of utilities designed to assist in the development of GPU-based algorithms, all accompanied by comprehensive documentation and a user-friendly interface.
Given these capabilities, we believe ByteQC will significantly advance quantum chemistry research and promote the wider adoption of GPU-accelerated methods within the scientific community.

Looking ahead, several new subpackages are already under development and will be released in the near future, including RPA and AFQMC\@.
We also plan to enhance the performance of existing subpackages with new implementations or algorithms, such as higher-order tensor contraction for exchange matrix construction, more efficient multi-GPU support for CCSD, and optimized implementations of CCSD(T).
To further improve efficiency, multi-node support is also being considered to enhance performance and strengthen competitiveness in the field.

\section*{Acknowledgments}

We would like to express our sincere gratitude to Dr.\ Hang Li and ByteDance Research for their invaluable support.
We also acknowledge Yaofeng Chen and Fan Yang for their contributions to the unreleased functionality of these packages, which will be integrated into the project in the future.
Additionally, we are grateful to Qiming Sun, Xiaojie Wu, and Hongzhou Ye for their insightful discussions, which significantly enhanced the project.

\FloatBarrier
\printbibliography
\end{document}